\documentclass{icrc2009}

\usepackage{graphicx}   
\usepackage{caption}    
\usepackage[font=footnotesize]{subfig} 
\usepackage{fixltx2e}
\usepackage{url}

\newcommand{\shorttitle}[1]%
{\markboth{Proceedings of the 31\MakeLowercase{$^{st}$} ICRC, {\L}\'{o}d\'{z} 2009}{#1} }
\newcommand{\etal}{\MakeLowercase{\textit{et al. }}} 

\begin{document}
\title{Evidence for a geomagnetic effect in the CODALEMA radio data}

\author{\IEEEauthorblockN{Beno\^{\i}t Revenu\IEEEauthorrefmark{1} for
the CODALEMA collaboration}

\IEEEauthorblockA{\IEEEauthorrefmark{1}SUBATECH, \'Ecole des mines de Nantes, CNRS/IN2P3, universit\'e de
Nantes,\\4 rue Alfred Kastler, 44307 Nantes Cedex 3 - FRANCE}}

\shorttitle{Revenu \etal CODALEMA geomagnetic effect}
\maketitle

\begin{abstract}
The CODALEMA experiment detects the cosmic air showers with an hybrid
detector composed of an array of scintillators and an array of antennas triggered by the particle array.
We will first describe the experiment and the data analysis then we will focus on the sky distribution of the events detected by the antennas which presents a strong asymmetry.
We will show that this asymmetry is well described and quantified at first order by an emission of the shower electric field given by the cross product of the shower axis with the geomagnetic field vector. The physical origin of this term is explained in another talk in this conference.
\end{abstract}

 
\section{Introduction}
\begin{figure}[!h]
  \centering
  \includegraphics[width=3in]{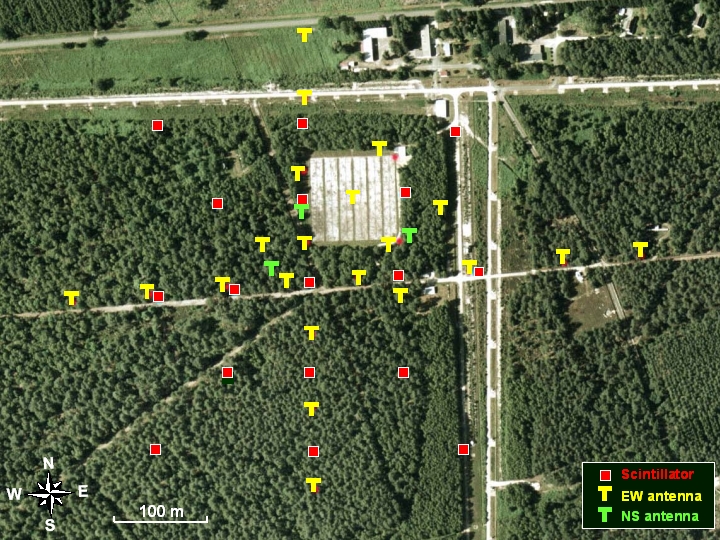}
  \caption{Schematic view of the CODALEMA experiment. The particle
array is represented by the squares. The radio array is indicated by
the letters ``T''.}
  \label{setup}
 \end{figure}

The radiodetection of cosmic air showers is studied by three
experiments: two in Europe (CODALEMA, France \cite{coda} and LOPES,
Germany \cite{lopes}) and one in Argentina, at the heart of the Pierre
Auger Observatory (see \cite{revenu,coppens}).
The main goal of the CODALEMA experiment, located at the Nan\c{c}ay radio
observatory, is the study of the electric field generated by the
charged particles of air showers created by the interaction of primary
cosmic rays in the atmosphere, at energies between $10^{16}$ and $10^{18}$~eV.
We are currently using two different
instruments: a particle array of 17 scintillators and a radio array of
24 antennas. The radio array is triggered by the particle array. The
estimation of the shower axis, core position and primary energy is
provided by the particle array and used as a reference by the
radio array. The detector setup is presented
in Fig.~\ref{setup}.
 The scintillators cover an area of $340\times
340$~m$^2$ and the distance between them is $85$~m. The antennas form
a cross with two arms of $600$~m length. The 14 antennas on the
North-South (NS) and East-West (EW) arms are used in this study and
these antennas measure the EW polarization of the incoming electric
field. We will also discuss in section~\ref{sec} the preliminary results we got with the 3
antennas measuring the NS polarization.

\section{The particle array}
Each particle detector is a plastic scintillator observed by two
photomultipliers with different high voltage gains in order to ensure
a dynamic range between $0.3$ and $3\,000$ Vertical Equivalent Muons (VEM).
The detectors are wired to a central shelter containing the power
supplies and the complete acquisition.
The start time and integrated signal of the ADC trace is computed for
each scintillator. The shower axis position, the core position and the
estimation of the number of particles reaching ground are computed
using a curved shower front and a Nishimura-Kamata-Greisen
function. Finally, the conversion of ground particle densities to
primary energy is obtained with the Constant Intensity Cut (CIC)
technique which defines a vertical equivalent shower size
$N_0=N(\theta=0^\circ,E)$. AIRES \cite{aires} simulations give the formula $E_0=2.14\times 10^{10}
~N_0^{0.9}$ with a resolution of $\Delta E/E\sim 30\%$ at $10^{17}$~eV
assuming protons as primaries. This resolution is mainly due to shower-to-shower
fluctuations and to the nature of the primary.
The reconstructed events are classified according to their core
position: internal events are those where the particle density is
larger for internal detectors than for external detectors lying on the
sides of the array. In this case, the core is contained inside the
array and we have a satisfactory accuracy for both shower
core and size. Energy estimation is meaningful only for internal
events; external ones have less reliable shower
parameters for core position and size but the axis position can be
used for further analysis.

The trigger condition requires the 5 central scintillators to have a signal
above $0.3$~VEM within a time window of $600$~ns. The average event rate is
around 8 events each hour.

\section{The radio array}

A single radio detector is constituted by a fat active dipolar antenna made
of two $60$~cm long, $10$~cm wide aluminium slats of $1$~mm thickness
separated by a $1$~cm gap and hold horizontally above the ground by a
$1$~m plastic mast.
The signal is preamplified by a dedicated, high input impedance, low
noise (1~nV.Hz$^{-1/2}$), $36$~dB amplifier with a $100$~kHz-$220$~MHz
bandwidth at $3$~dB. To avoid intermodulation due to a $2$~GW local emitter, the
signal is high-pass filtered ($20$~dB at 162~kHz). The 14 antennas
are triggered by the particle array; the digitization is done by a fast 12 bits ADC running at
$1$~GHz with a memory depth of $2\,560$ points (corresponding to $2.5~\mu$s).

\begin{figure}[!h]
  \centering
  \includegraphics[width=3.2in]{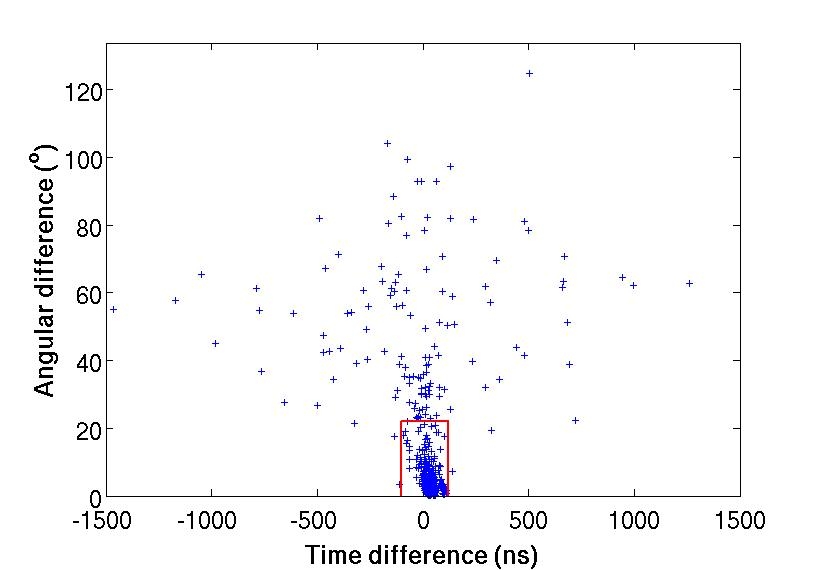}
  \caption{Selection of the actual cosmic events detected by the
radio array using the angular difference and the time difference between the
two independent reconstructions; the box indicates the criteria for
coincidences: angular difference smaller than $20^\circ$ and time
difference in the interval $[-100,100]$~ns.}
  \label{tree}
 \end{figure}

The offline analysis first filters the signal in the $23-83$~MHz band
after correction for the cable frequency response. We use the galactic
radio background to adjust the relative gains. In this frequency band,
the pulse corresponding to the transient event is found using the Linear
Prediction Method~\cite{lpc}. Then, the pulse is associated to an absolute
time and a signal value. From the time information of the tagged antennas we can
reconstruct an arrival direction by simple triangulation. Using the time information of
the two arrays, we have two independent measurements of the
shower axis and impact time which can be compared in order to identify
unambiguously cosmic shower seen by the radio array. The set of events seen by the
radio array is consituted of events having an angular difference
smaller than $20^\circ$ and a time difference in the interval
$[-100,100]$~ns, as illustrated in Fig.~\ref{tree}.

\section{Radio detection efficiency}

\begin{figure}[!h]
  \centering
  \includegraphics[width=3.4in]{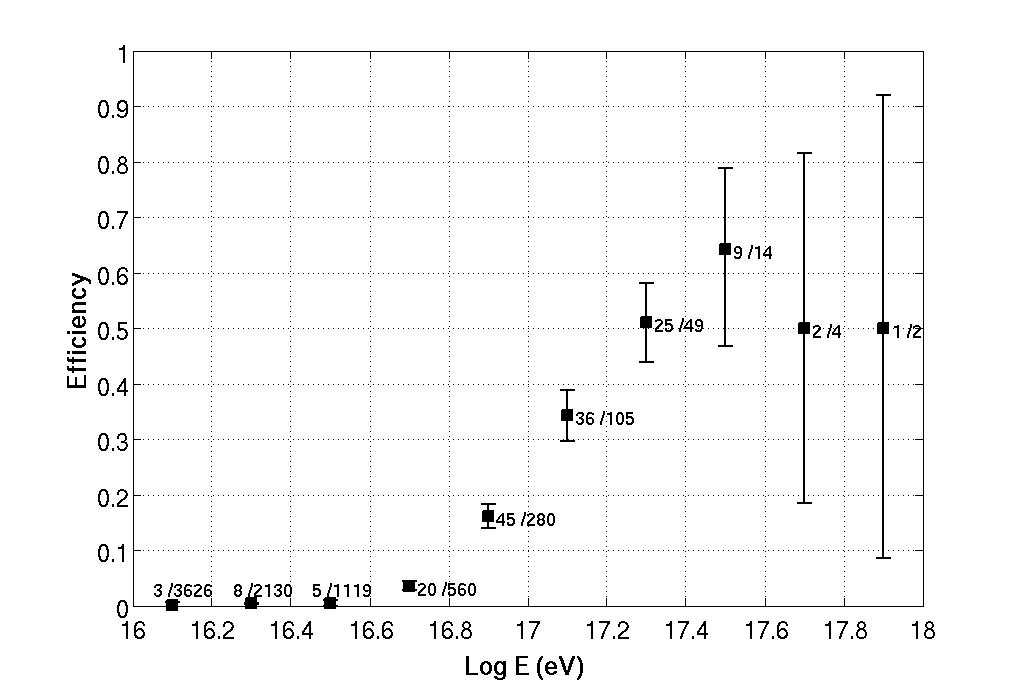}
  \caption{Efficiency of  the radio array defined as the ratio of
the number of radio events
to the number particle events in a given energy bin.}
  \label{eff}
 \end{figure}

The data set presented in this analysis has been recorded between November $27^\mathrm{th}~2006$
and March $20^\mathrm{th}~2008$ corresponding to $355$ effective days
of stable data acquisition. The numbers of internal events in this period are
summarized in Table~\ref{tab}.
\begin{table}[!h]
  \caption{Number of events.}
  \label{tab}
  \centering
  \begin{tabular}{|c|c|c|c|}
\hline
events & scintillators & antennas & coincidences \\ \hline
reconstructed & $61\,517$ & $750$ & $619$ \\ \hline
internal & $28\,128$ & $195$ & $157$ \\ \hline
$E>10^{16}$~eV & $7\,889$ & $169$ & $154$ \\ \hline
$E>5\times 10^{16}$~eV & $692$ & $134$ & $129$ \\ \hline
\end{tabular}
\end{table}
The energy threshold\footnote{The threshold energy is defined as the energy for which we
detect $80\%$ of the events.} of the radio array appears to be around
$10^{17}$~eV and that of the particle array around $10^{15}$~eV. The efficiency of the radio array is defined as the ratio of
the number of radio events
to the number of particle events in a given energy range. This
efficiency is presented in Fig.~\ref{eff}. It is increasing regularly
above $3\times 10^{16}$~eV and reaches $50\%$ at $2\times 10^{17}$~eV.
Taking into account the properties of the electric field generated by
the geomagnetic field permits to be compatible with a full efficiency
at high energy, as argued in section~\ref{sec}.

\section{Asymmetry in the radio events}
\begin{figure}[!h]
  \centering
  \includegraphics[width=3.4in]{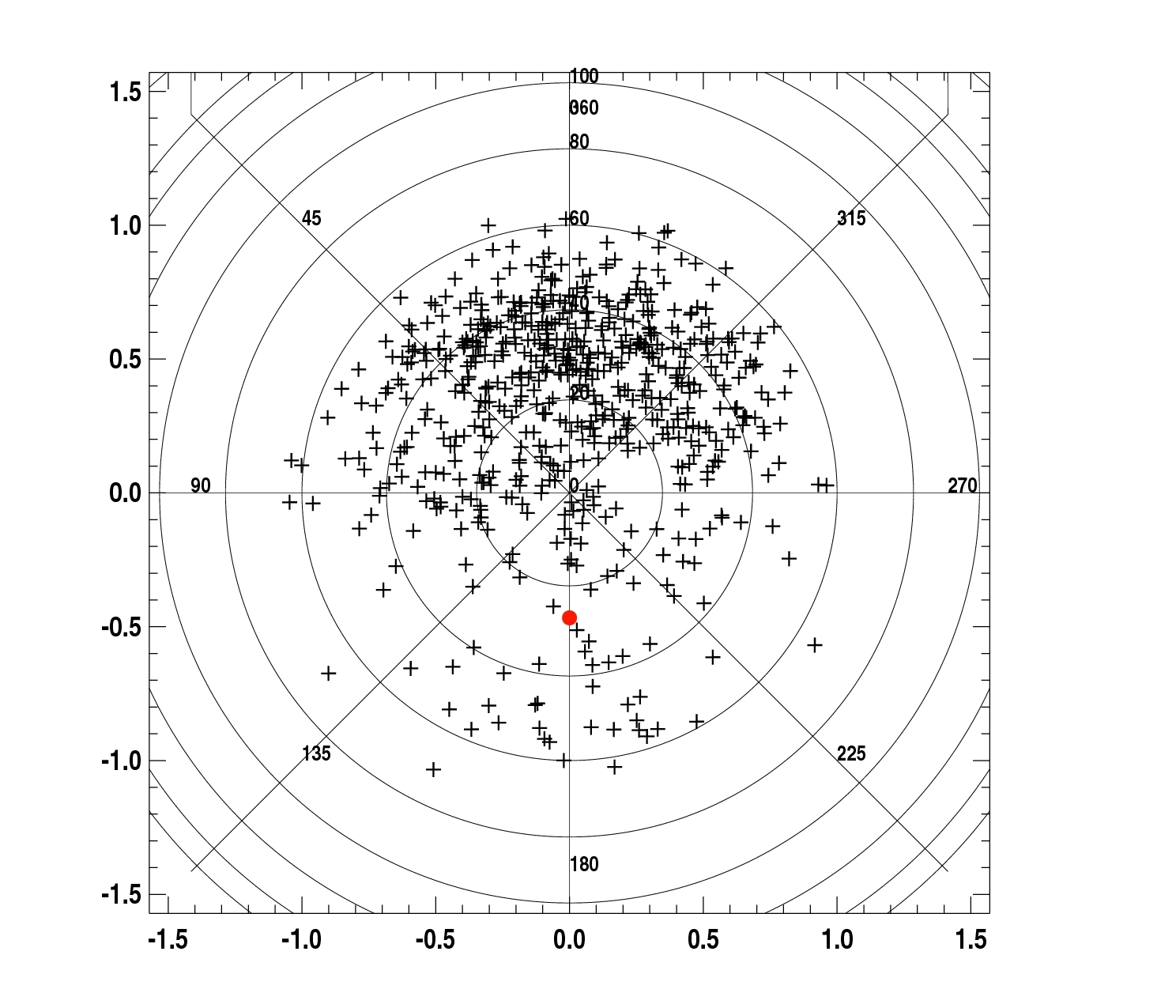}
  \caption{Sky map of observed radio events. The zenith is at the
center and the azimuths are: North (top, $0^\circ$), West
(left, $90^\circ$), South (bottom, $180^\circ$), East (right,
$270^\circ$). The geomagnetic field is indicated by the dot.}
  \label{sky}
 \end{figure}

We first focus on the list of all events seen in coincidence between
the particle array and the radio array, including external events for
which the axis position only is known.
The projection on a sky map of the arrival directions of these events is presented in Fig.~\ref{sky}.
We clearly
observe a strong relative excess of events coming from the North. The
ratio of the number of events coming from the South (ie with azimuth
$90^\circ<\phi<270^\circ$) to the total number of events is
$n_\mathrm{South}/n_\mathrm{tot}=109/619=0.17\pm 0.02$. This asymmetry
is not expected a priori because:
\begin{itemize}
\item the geometrical setup of the CODALEMA experiment is symmetric
and presents no instrumental bias;
\item the azimuthal distribution of the particle array (remind that
it also provides the trigger of the radio array) is compatible with a
flat distribution.
\end{itemize}
We compared the observed sky map to the sky map one could naively have
expected, namely a symmetrized version of the
coverage map obtained from the particle zenithal distribution with a
flat azimuthal distribution: the discrepancy reaches 16 standard deviations. We
verified that the asymmetry level is independent of the data set:
7 independent time ordered sets of events give similar values of the
ratio $n_\mathrm{South}/n_\mathrm{tot}$ (around $0.17$).

\begin{figure}[!h]
  \centering
  \includegraphics[width=3in]{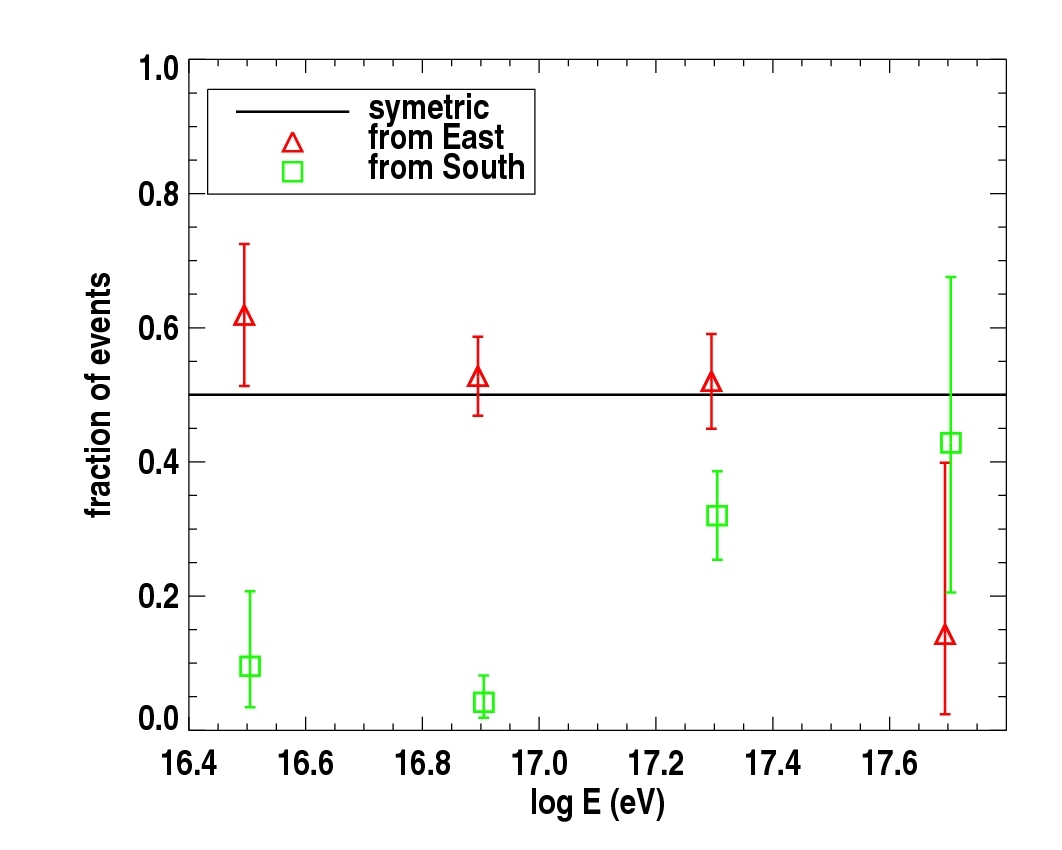}
  \caption{Evolution of the ratio $n_\mathrm{South}/n_\mathrm{tot}$ as
a function of energy (energy bins are independent). A ratio of 0.5
indicates symmetry. At high energy, the
observed radio sky map become compatible with a symmetric map in both
NS/EW sectors.}
  \label{asym2}
 \end{figure}

If we restrict the data set to internal events (with an
estimation of the primary energy), we observe that when the energy
increases, the asymmetry level becomes compatible with the symmetric
value of $n_\mathrm{South}/n_\mathrm{tot}=0.5$, as shown in
Fig.~\ref{asym2}. The ratio $n_\mathrm{East}/n_\mathrm{tot}$ is always
compatible with $0.5$.
We can therefore conclude that the asymmetry observed by the radio
array is due to an energy threshold effect.

\section{Semi-empirical model}\label{sec}
\begin{figure}[!h]
  \centering
  \includegraphics[width=3in]{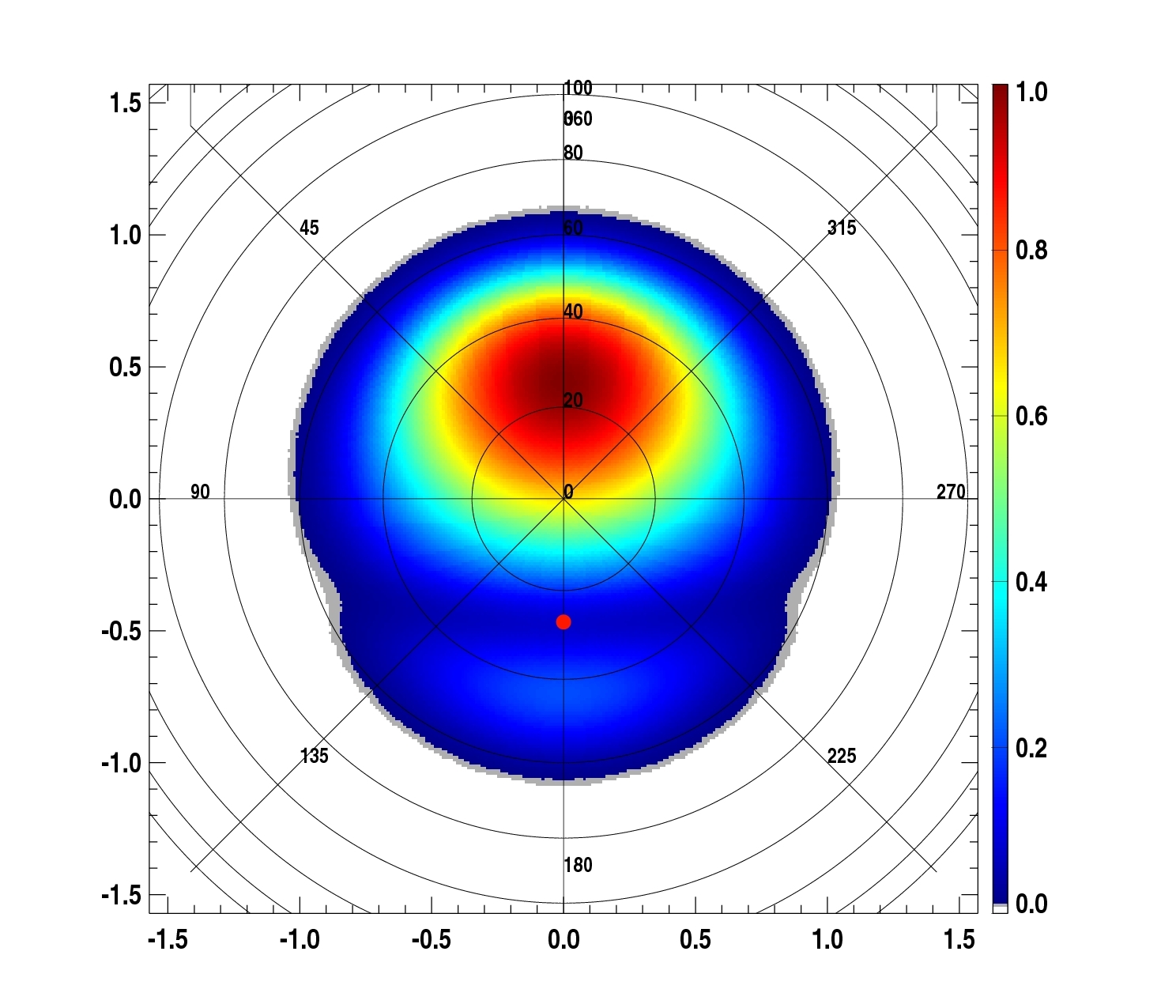}
  \caption{Expected EW density map according to the
semi-empirical model. See text for details.}
  \label{toym}
 \end{figure}

The detected asymmetry is clear and unambiguous. In
order to understand its origin, we can think about the effect of the
geomagnetic field on the charged secondary particles of the shower. 
 The associated Lorentz force acting on
these particles is responsible for the emission of the electric
field. We can therefore expect a rough dependence of the global
macroscopic electric field on a Lorentz term such as $\vec{E}\propto \vec{v}\times
\vec{B}$ where $\vec{v}$ is the shower axis direction and $\vec{B}$ the
geomagnetic field.
The EW-oriented dipolar antennas will be sentitive to the EW component of this electric field $(\vec{v}\times\vec{B})_\mathrm{EW}$. 
To build the expected event map, we must also
take into account the zenithal distribution of the particle array
which triggers the radio array. We used for this the
following parametrization:
\begin{displaymath}
\frac{\mathrm{d}N}{\mathrm{d}\theta}=(a+b\,\theta)\frac{\cos
\theta\sin\theta}{1+\exp((\theta-\theta_0)/\theta_1)},
\end{displaymath}
with $a=44.96$, $b=0.57$, $\theta_0=49.18^\circ$,
$\theta_1=5.14^\circ$, computed using the internal events above
$10^{17}$~eV.
The resulting expected events density
sky map taking into account all these ingredients is presented in
Fig.~\ref{toym}, assuming that the number of events is
proportionnal to the amplitude of the electric field.
An efficient way to test the validity of this semi-empirical model is to
compute an ensemble average of the angular distributions of a large
number of realisations of $N$ simulated events
following the expected density map, where $N$ is the actual number of
detected events in a given polarization: $N=619$ for the EW
polarization and $N=100$
for the NS polarization.
\begin{figure*}[!h]
  \centerline{\subfloat[Azimuthal distribution for the EW polarization.]{\includegraphics[width=3.2in]{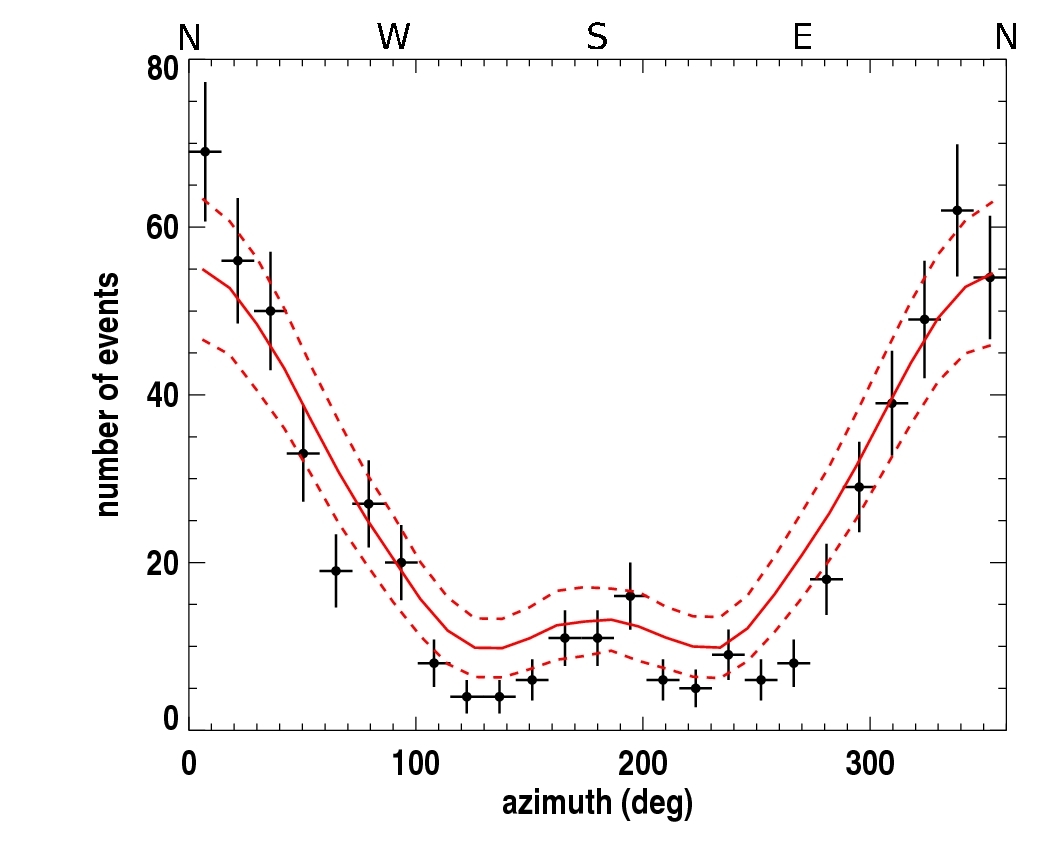}\label{distp}}
\hfil
\subfloat[Azimuthal distribution for the NS polarization.]{\includegraphics[width=3.2in]{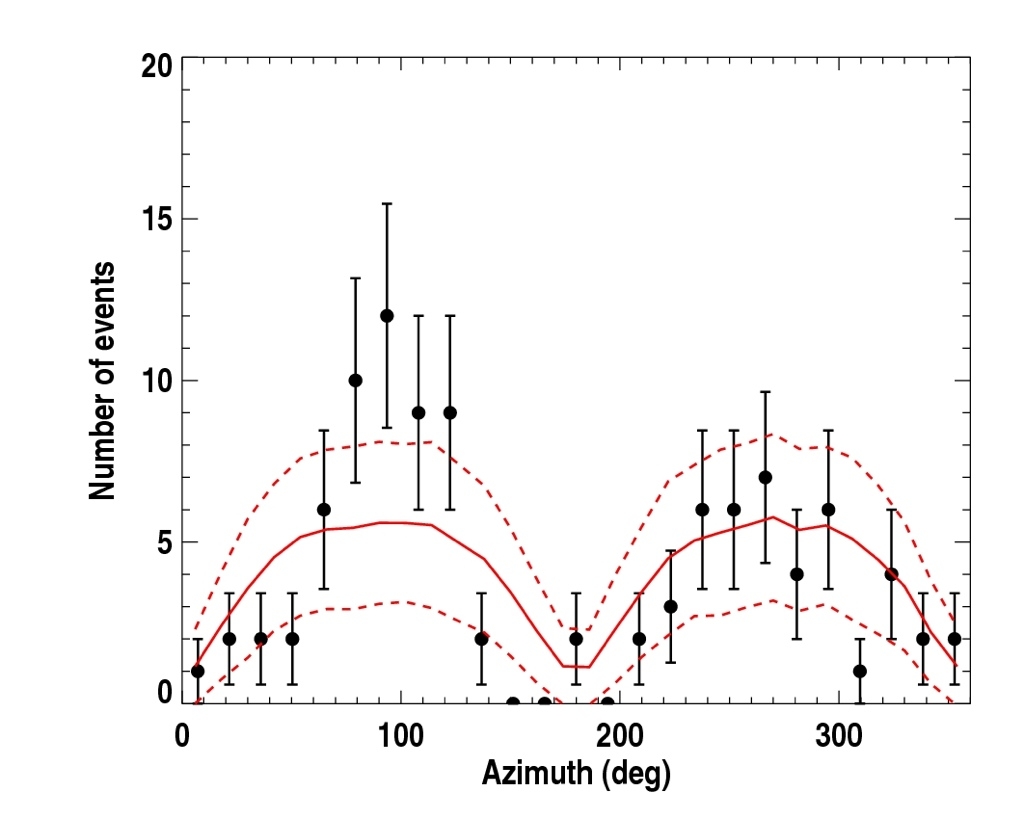}\label{distNSp}}}
\caption{Comparison of the expected azimuthal distributions to the
observed azimuthal distributions (crosses) in the EW polarization
(Fig.~(a) and~(b)) and in the
NS polarization (Fig.~(c) and~(d)). The lines correspond to the
simulated azimuthal
distributions obtained after $1\,000$
realisations of $619$ events and $100$ events for the EW and NS
polarizations respectively with the associated $\pm 1\sigma$ error bands.}
 \end{figure*}
 In our case, for the EW and
NS polarization, the zenithal observed and simulated distributions are
in very good agreement; Figures~\ref{distp} and~\ref{distNSp}
present the results for the azimuthal distributions.
The semi-empirical model reproduces correctly the main features of the
observed azimuthal distributions: for the EW polarization we have a maximum towards the North, a local maximum
towards the South and minima in the East and West directions; for the
NS polarization we have events in both East and West hemisphere and no
events coming from the North and the South, in agreement with the
semi-analytical model.
The semi-analytical model describes remarkably well the angular
distributions in both EW and NS polarizations. The study of the
polarity of the electric field can be found in \cite{coda09} and in
another contribution to this conference \cite{colas}.

The model permits to understand the shape of the sky distribution
but can also be used to check the radio detection efficiency as a
function of the geometrical factor $\vec{v}\times
\vec{B}$. For the EW polarization (for which we have higher statistics), when plotting the ratio of the number of radio
detected events to the total number of events as a function of $(\vec{v}\times
\vec{B})_\mathrm{EW}/|vB|$, we find a linear
increase from $0$ to $1$ which validates our assumption that the
number of detected events is directly linked (nearly proportional) to
the electric field value. We observe that in order to be detected,
events with a low value of the Lorentz force EW component (as it is
the case for events coming for
instance from East or West) must have a higher energy than events with
high values, coming from North. At the highest energies, all events
are detectable independently of their arrival direction and the radio
efficiency will become independent of the Lorentz force component. In
Fig.~\ref{eff}, the radio efficiency was plotted as a function of the
energy given by the particle array. If we rescale this energy
according to the semi-analytical model $E'=E\,.\,(\vec{v}\times
\vec{B})_\mathrm{EW}/|vB|$,
we obtain an efficiency which saturates at $100\%$ above the
threshold energy $E_\mathrm{th}=10^{17}$~eV (we had a saturation value of $50\%$ in Fig.~\ref{eff}).

\section{Conclusion}

Around the threshold
energy $E_\mathrm{th}=10^{17}$~eV of the radio array, we detect cosmic showers with an important level
of asymmetry in both EW and NS polarizations. At higher energy, the azimuthal
distributions become compatible with symmetric distributions as
expected. The measured electric field is well described by the average
action of the Lorentz force resulting into a macroscopic field of the form
$\vec{E}=\vec{v}\times\vec{B}$. This model reproduces the details of
the azimuthal distributions and a rescaling of the energy by the
geometrical factor of the model permits to reach full efficiency above
the threshold energy. The resulting parametrization is not compatible
with the one obtained by the LOPES experiment \cite{horn}.
In the near future (end 2009), we plan to install the next generation
of radio detectors at the Nan\c{c}ay Observatory and at the Pierre
Auger Observatory.


\end{document}